\documentstyle{article}
\begin{document}
\hoffset -1.5cm
\renewcommand{\baselinestretch}{1.2}
\title{\Large\bf
MASS SCALES AND STABILITY OF THE PROTON IN [SU(6)]$^{3}\times$Z$_{3}$}
\author{\large
William A. Ponce$^{1,2}$, Arnulfo Zepeda$^2$ \\
and Juan B. Fl\'orez$^{2,3}$ \\
\normalsize  1-Departamento de F\'\i sica, Universidad de Antioquia\\
\normalsize A.A. 1226, Medell\'\i n, Colombia \\
\normalsize  2-Departamento de F\'\i sica, Centro de Investigaci\'on y
estudios avanzados del I.P.N. \\
\normalsize Apartado Postal 14-740, 07000 M\'exico D.F., M\'exico \\
\normalsize 3-Departamento de F\'\i sica, Universidad de Nari\~no \\
\normalsize A.A. 1175 San Juan de Pasto, Colombia}

\setlength{\hsize}{15cm}

\maketitle
\pagebreak
\large

\vspace{2cm}

\begin{center}
{\bf ABSTRACT}
\end{center}
\large
\parbox{15cm}%
{We proof that the proton is stable in the left-right symmetric gauge
model [SU(6)]$^3\times$Z$_3$,
which unifies nongravitational forces with flavors,
broken spontaneously by a minimal set of Higgs Fields and Vacuum
Expectation Values down to SU(3)$_{c}\otimes$U(1)$_{EM}$\hspace{0.2cm} . We
also
    compute the evolution
of the gauge coupling constants and show how agreement with precision
data can be obtained.}

\pagebreak

\large
\section{Introduction}
Recently, we have proposed \cite{gmpz,pz} a grand unification model (GUM)
of forces and flavors based on the gauge group G=[SU(6)]$^3\times$Z$_3$.
Our aim has been to provide some clues for the explanation of the intriguing
fermion mass spectrum and mixing parameters (since G includes the so
called horizontal interactions, it has predictions for the
masses of some elementary fermions). Comparing our model with related attempts
made in the context of GUMs with simple groups without discreet symmetries
(such as \cite{zee} SO(18) and \cite{bars} E$_8$), we find it simpler,
more economical and more elegant than their competitors. Not only the model
described in  \cite{gmpz,pz} does not contain mirror fermion fields, but
it has fewer gauge fields, fermion fields and Higgs fields than the other two
mentioned models.

As it is clear from Refs.  \cite{gmpz} and  \cite{pz}, the fermion
content of this model includes in a
single irreducible representation of G the three families of known fermions,
each family being defined by the dynamics of the
SU(3)$_c\otimes$SU(2)$_L\otimes$SU(2)$_R\otimes$U(1)$_{Y_{(B-L)}}$
gauge group. This last group is the
left-right symmetric (LRS) extension of the
SU(3)$_{c}\otimes$SU(2)$_{L}\otimes$
   U(1)$_{Y}$ standard model (SM).

The rest of this communication is as follows: In section two we
review the model, in section three we do the renormalization group
equation analysis, and in section four we demonstrate the stability
of the proton in the context of our model.

\vspace {3mm}
\section{The model}
The gauge group is given by
G=SU(6)$_{L}\otimes$SU(6)$_{C}\otimes$SU(6)$_{R}\times$Z$_{3}$
 \cite{gmpz}. SU(6)$_C$ is a vector like group which includes three
hadronic and three leptonic
colors. SU(6)$_C$ includes as a subgroup the
SU(3)$_c\otimes$U(1)$_{Y_{(B-L)}}$ of the LRS model.
SU(6)$_{L}\otimes$SU(6)$_{R}$ is
the left-right symmetric flavor group which includes the
SU(2)$_{L}\otimes$SU(2)$_{R}$ gauge group of the LRS model, and
SU(3)$_{HL}\otimes$SU(3)$_{HR}$ is the horizontal gauge group in G.

\subsection{The Gauge and Fermion Fields}
The 105 gauge fields (GF) and the 108 Weyl Fermions in G are explicitly
depicted in Ref. \cite{gmpz,pz}. Let us describe here some of them:

The 105 GF can be divided in two sets, 70 of them belonging to
SU(6)$_L\otimes$SU(6)$_R$ and 35 associated with SU(6)$_C$. The first
set includes W$_L^\pm$ and Z$^0_L$, the GF of the known weak interactions,
plus the GF associated with the postulated right weak interaction, plus
the GF of the horizontal interactions, etc.. All of them are SU(3)$_c$
singlets and have electrical charges 0 or $\pm$ 1. The second set includes
the 8 gluon fields of SU(3)$_c$; nine leptoquark GF, X$_i$, Y$_i$, and
Z$_i$, $i=1,2,3$ with electrical charges $-2/3$, 1/3, and $-2/3$ respectively;
other 9 leptoquark GF charge conjugated to the previous ones;
six diquark GF, P$_a^\pm$,
P$^0$ and $\stackrel{\sim}{P}^0$, $a=1,2$, with electrical charges as
indicated, and three GF associated with diagonal generators in SU(6)$_C$,
where B$_{Y_{(B-L)}}$ the GF associated with the U(1)$_{_{Y(B-L)}}$
factor in the LRS model is one of them.

The ordinary (known) fermions for the model are included in \\
$\psi(108)_L =$Z$_3\psi(6,1,\bar{6})\equiv\psi(6,1,\bar{6})_{L} +
\psi(1,\bar{6},6)_{L} + \psi(\bar{6},6,1)_{L}$,
with the following particle content:

\begin{equation}
\psi(\bar{6},6,1)_{L} = \left( \begin{array}{cccccc}
{\rm d}_{x}^{-1/3} & {\rm d}_{y}^{-1/3} & {\rm d}_{z}^{-1/3} & E_{1}^{-} &
L_{1}^{0} & T_{1}^{-}\\
{\rm u}_{x}^{2/3} & {\rm u}_{y}^{2/3} & {\rm u}_{z}^{2/3} & E_{1}^{0} &
L_{1}^{+} & T_{1}^{0}\\
{\rm s}_{x}^{-1/3} & {\rm s}_{y}^{-1/3} & {\rm s}_{z}^{-1/3} & E_{2}^{-} &
L_{2}^{0} & T_{2}^{-}\\
{\rm c}_{x}^{2/3} & {\rm c}_{y}^{2/3} & {\rm c}_{z}^{2/3} & E_{2}^{0} &
L_{2}^{+} & T_{2}^{0}\\
{\rm b}_{x}^{-1/3} & {\rm b}_{y}^{-1/3} & {\rm b}_{z}^{-1/3} & E_{3}^{-} &
L_{3}^{0} & T_{3}^{-}\\
{\rm t}_{x}^{2/3} & {\rm t}_{y}^{2/3} & {\rm t}_{z}^{2/3} & E_{3}^{0} &
L_{3}^{+} & T_{3}^{0}
\end{array}
\right) _{L}\equiv\psi^{\alpha}_{a},
\end{equation}

\noindent
where the rows (columns) represent color (flavor) degrees of
freedom; E$_i^{-,0}$, L$_i^{+,0}$, and T$_i^{-,0}$ $i=1,2,3$ stand
for leptonic fields with electrical charges as indicated, and
d,u,s,c,b,and t stand for the corresponding quark fields
eigenstates of G, but not mass eigenstates \cite{gmpz}; the
subindices $x,y,z$ in the quark fields refers to SU(3)$_c$ color
indices. $\psi(1,\bar{6},6)_L\equiv\psi^A_\alpha$ includes the 36
fields charge conjugated to the fields in $\psi(\bar{6},6,1)_L$, and
$\psi(6,1,\bar{6})_{L}\equiv\psi^a_A$
represents  36 exotic Weyl leptons, 9 with
positive electric charges, 9 with negative (the charge
conjugated to the positive ones) and 18 are
neutrals. As it is clear, we are using a,b,...as SU(6)$_L$ tensor indices;
A,B,... as SU(6)$_R$ tensor indices, and $\alpha,\beta,..$ as
SU(6)$_C$ tensor indices.

\subsection{The Higgs Fields}
The analysis done in Ref. \cite{pz} shows that
the most economical set of Higgs Fields (HF) and
Vacuum Expectation Values (VEVs) which breaks the symmetry
from G down to SU(3)$_{c}\otimes$U(1)$_{EM}$\hspace{0.2cm} and at the same time
   produces what we called
the {\bf modified horizontal survival hypothesis} is
formed by:

\begin{equation}
\phi_{1}=\phi(675)=
\phi_{1,[a,b]}^{[A,B]}+\phi^{[\alpha,\beta]}_{1,[A,B]}
+\phi^{[a,b]}_{1,[\alpha,\beta]}       \label{fi1}
\end{equation}

\noindent
with VEVs in the directions [a,b]=[1,6]=$-$[2,5]=$-$[3,4], [A,B] similar to
[a,b] and [$\alpha,\beta$]=[5,6];

\begin{equation}
\phi_{2}=\phi(1323)=
\phi_{2,\{a,b\}}^{\{A,B\}}+\phi^{\{\alpha,\beta\}}_{2,\{A,B\}}
+\phi^{\{a,b\}}_{2,\{\alpha,\beta\}}       \label{fi2}
\end{equation}

\noindent
with VEVs in the directions \{a,b\}=\{1,4\}=$-$\{2,3\}, \{A,B\} similar to
\{a,b\} and \{$\alpha,\beta$\}=\{4,5\};

\begin{equation}
\phi_{3}=\phi^\prime(675)=
\phi_{3,[a,b]}^{[A,B]}+\phi^{[\alpha,\beta]}_{3,[A,B]}
+\phi^{[a,b]}_{3,[\alpha,\beta]}       \label{fi3}
\end{equation}

\noindent
with VEVs such that $\langle\phi_{3,[a,b]}^{[A,B]}\rangle=
\langle\phi^{[a,b]}_{3,[\alpha,\beta]}\rangle=0$, and
$\langle\phi^{[\alpha,\beta]=[4,6]}_{3,[A,B]=[4,6]}\rangle=$M$_R$;

\begin{equation}
\phi_4=\phi(108)=\phi^A_{4,\alpha}+\phi^\alpha_{4,a}+\phi^a_{4,A}.
\label{eq-fi4}
\end{equation}

\noindent
with VEVs such that
$\langle\phi_\alpha^A\rangle=\langle\phi_a^\alpha\rangle=0$ and
$\langle\phi_A^a\rangle=$M$_L$, with
values different from zero only in the directions
$\langle\phi_2^2\rangle=
\langle\phi_4^2\rangle=
\langle\phi_6^2\rangle=
\langle\phi_2^4\rangle=
\langle\phi_4^4\rangle=
\langle\phi_6^4\rangle=
\langle\phi_2^6\rangle=
\langle\phi_4^6\rangle=
\langle\phi_6^6\rangle=$M$_Z\sim 10^2$ GeVs.

In (\ref{fi1}), (\ref{fi2}), and (\ref{fi3}), [.,.]
and \{.,.\} stand for the commutator and anticommutator
respectively of the indices inside the brackets.
The mass hierarchy suggested in Ref. \cite{pz} is
$\langle\phi_3\rangle > \langle\phi_1\rangle\simeq\langle\phi_2\rangle
\gg M_Z\sim 10^2$ GeVs. Since at this stage of our analysis we are
not interested in studying CP violation in the context of this
model, we will assume throughout the paper that
$\langle\phi_i\rangle$, $i=1,2,3,4$, are real numbers.

According to the analysis presented in Ref.  \cite{pz},
$\langle\phi_1\rangle+\langle \phi_2\rangle$ breaks G down to the LRS
gauge group; $\langle\phi_3\rangle$ alone breaks G down to
SU(6)$_L\otimes$SU(4)$_C\otimes$U(1)$_Y\otimes$G$_R$, where U(1)$_Y$ is
the same abelian factor of the SM and SU(4)$_C\supset$SU(3)$_c$ but
U(1)$_{Y_{(B-L)}}$ is not a subgroup of SU(4)$_C$ neither SU(2)$_R$
of the LRS model is a subgroup of G$_R$.
Finally, $\langle\phi_1\rangle +\langle \phi_2\rangle
+\langle\phi_3\rangle$ breaks G down to
SU(3)$_{c}\otimes$SU(2)$_{L}\otimes$U(1)
   $_{Y}$,
the local gauge group of the SM.

Let us emphasize that prior to symmetry breaking the model is both, left-right
symmetric and iso-spin symmetric. It is the choice of VEVs for
$\langle\phi_4\rangle$ which resulted in an spontaneous breaking of the
iso-spin symmetry and in a flavor democratic mass matrix for the Up quarks.
The iso-spin breaking was introduced by hand, by demanding that
the only VEVs of $\langle\phi_4\rangle\neq 0$ are the ones stated above. This
particular choice of VEVs is what produces the {\bf modified horizontal
survival hypothesis} \cite{gmpz,pz}
(the only ordinary particle with a tree level mass different from zero is
the t quark). On the other hand, the flavor democratic form of the mass
matrices for the quark sectors  resulted as a consequence of minimizing
the Higgs
field potential for $\phi_4$ alone, and it is hoped that a
careful study of the entire Higgs fields potential provides a basis for
the iso-spin breaking too.

\section{The Renormalization Group Equation Analysis}
For the renormalization group equation (RGE) analysis which follows, we
implement our model with the working conditions known as ``{\it the
survival hypothesis"} \cite{sh} and ``{\it the extended survival
hypothesis"}  \cite{esh}.

\noindent
1-The survival hypothesis claims that \cite{sh} at each energy scale, the only
fermion fields which are relevant are those belonging to chiral
representations of the unbroken symmetries.

\noindent
2-The extended survival hypothesis claims that \cite{esh} at each energy
scale the only scalars which are relevant are those that develop VEVs
at that scale and at lower mass scales.

In the context of our model we can show that both hypothesis follow
mandatory by a wise selection of VEVs for the scalar fields, and the
inclusion of appropriate terms in the scalar potential and Yukawa Lagrangian.

Now, for the proper implementation of the survival hypothesis,
the Higgs scalars which develop VEVs must
couple to $\psi(108)_L$ via Yukawa type terms. Therefore they must be
of the form Z$_3\phi(n,1,m)$ where n,m=$6,\bar{6},15,\overline{15},21,
\overline{21}$. (Higgs fields of the form Z$_3\phi(6,35,\bar{6})$, and Higgs
fields which do not develop VEVs are excluded from our model for economic
reasons  \cite{pz}.)

\subsection{The RGE analysis with two mass scales.}
For the two mass scale symmetry breaking pattern \\
G$\stackrel{M}{\longrightarrow}$SU(3)$_c\otimes$SU(2)$_L\otimes
$U(1)$_Y\stackrel{M_Z}{\longrightarrow}$SU(3)$_c\otimes$U(1)$_{EM}$\\
where M=$\langle\phi_1\rangle+\langle\phi_2\rangle+\langle\phi_3\rangle$ and
M$_Z=\langle\phi_4\rangle$, the one loop running coupling constants of
the standard model satisfy the relationships

\begin{equation}
\alpha_i^{-1}(M_Z)=\alpha_i^{-1}(M)-b_i^0{\rm ln}(M/M_Z)
\label{eq-a1}
\end{equation}
\noindent
where $\alpha_i=$g$_i^2/4\pi$, $i=1,2,3$ refers to U(1)$_Y$, SU(2)$_L$
and SU(3)$_c$ respectively, and the beta functions $b_i$ are given by

\begin{equation}
b_i^{\kappa}=\{\frac{11}{3}C_i^{\kappa}(vectors)-
\frac{2}{3}C_i^{\kappa}(Weyl-fermions)-
\frac{1}{6}C_i^{\kappa}(scalars)\}/4\pi
\label{eq-a2}
\end{equation}
\noindent
with $C_i^{\kappa}(...)$ the index of the representation to which the
(...) particles are assigned. For a complex field, the value
$C_i^{\kappa}(scalars)$ should be doubled.
With the normalization of the generators in G such that
$\alpha_1(M)=\alpha_2(M)=\alpha_3(M)$, the relationship

\begin{equation}
\alpha_{EM}=\frac{1}{3}\alpha_2{\rm sin}^2\theta_W
=\frac{3}{14}{\rm cos}^2\theta_W,
\label{eq-a3}
\end{equation}

\noindent
where $\theta_W$ is the weak mixing angle, is valid at all energy scales.
This last equation implies also that at all energies

\begin{equation}
3\alpha^{-1}_{EM}=14\alpha_1^{-1}+9\alpha_2^{-1}.
\label{eq-a444}
\end{equation}

\noindent
Equations (~\ref{eq-a1}), (~\ref{eq-a3}) and (~\ref{eq-a444})
give straightforward

\begin{equation}
\frac{3}{23}\alpha^{-1}_{EM}(M_Z)=\alpha_3^{-1}(M_Z)+
(b_3^0-\frac{14}{23}b_1^0-\frac{9}{23}b_2^0){\rm ln}(M/M_Z)
\label{eq-a6}
\end{equation}
\noindent
and
\begin{eqnarray}
{\rm sin}^2\theta_W(M_Z)&=&3\alpha_{EM}(M_Z)\alpha_2^{-1}(M_Z) \nonumber \\
&=&3\alpha_{EM}(M_Z)[\alpha_3^{-1}(M_Z)+(b_3^0-b_2^0){\rm ln}(M/M_Z)]
\label{eq-a7}
\end{eqnarray}
\noindent
where $b_3^0=(11-4)/2\pi$, $b_2^0=[\frac{22}{9}-\frac{4}{3}(3-n^0_2)
-\frac{N_H}{18}]/2\pi$ and $b_1^0=-[\frac{4}{3}(3-n^0_1)
+\frac{N_H}{28}]/2\pi$,
$N_H=9$ is the number of low energy Higgs fields doublets in
$\langle\phi_4\rangle$, and $n^0_2=2,n^0_1=27/14$ are two values related
to the number of fermion fields which decouple from $\psi(108)_L$ according
to the survival hypothesis and the Appelquist-Carrazone theorem  \cite{apel}
($n^0_1=n^0_2=0$ when all the fermion fields in $\psi(108)_L$ contribute to
$b^0_{1,2}$).

Plugging in the last two equations the experimental values \cite{amaldi}
sin$^2\theta_W(M_Z)=0.233$, $\alpha^{-1}_{EM}(M_Z)=127.9$, and
$\alpha_3(M_Z)=0.122$ we get from Eq.(~\ref{eq-a6}) ln(M/M$_Z)=6.3$,
and from Eq.(~\ref{eq-a7}) ln(M/M$_Z)=1.1$ which are widely incompatible
and inconsistent solutions.
So, the model with only two mass scales is excluded by experimental
results (this same conclusion was reached in a different way in
Ref. \cite{pz}).

\subsection{The RGE analysis with three mass scales}
For the three mass scale symmetry breaking pattern\\
G$\stackrel{M}{\longrightarrow}$G$_L\otimes$G$_C\otimes$G$_R\otimes ...
\stackrel{M_H}{\longrightarrow}
$SU(3)$_c\otimes$SU(2)$_L\otimes$U(1)$_Y\stackrel{M_Z}
{\longrightarrow}$SU(3)$_c\otimes$ U(1)$_{EM}$ \\
where $M>>M_H>>M_Z=\langle\phi_4\rangle$,
the one loop running coupling constants of the standard model satisfy now
\begin{equation}
\alpha_i^{-1}(M_Z)=\alpha_i^{-1}(M)-b_i^0{\rm ln}(M_H/M_Z)-
b_i^1{\rm ln}(M/M_H),
\end{equation}
\noindent
Now the algebra gives
\begin{equation}
\frac{3}{23}\alpha^{-1}_{EM}(M_Z)=\alpha_3^{-1}(M_Z)
+(b^0_3-\frac{14}{23}b^0_1-\frac{9}{23}b^0_2){\rm ln}(\frac{M_H}{M_Z})
+(b^1_C-\frac{14}{23}b_Y^1-\frac{9}{23}b^1_L){\rm ln}(\frac{M}{M_H})
\label{w}
\end{equation}
\noindent
and
\begin{equation}
{\rm sin}\theta_W(M_L)=3\alpha_{EM}(M_Z)[\alpha_3^{-1}(M_Z)
+(b_3^0-b^0_2){\rm ln}(\frac{M_H}{M_Z})
+(b_C^1-b^1_L){\rm ln}(\frac{M}{M_H}),
\label{ww}
\end{equation}
\noindent
where $b_i^0,i=1,2,3$ are the same as in Sec. 3.1, but $b^1_i,i=C,Y,L$ depend
upon the structure of the subgroup
$G_L\otimes G_C\otimes G_R\otimes...\equiv G_H$
as mentioned in Ref. \cite{pz}.

Equation (\ref{ww}) is very restrictive because its first entry of the
right-hand side has an experimental value of 0.192 which excludes the
possibility of having\\
G$_H=$SU(3)$_C\otimes$SU(2)$_L\otimes$SU(2)$_R\otimes$U(1)$_{Y(B-L)}$
(which shows up for $M=\langle\phi_1+\phi_2\rangle)$
because if so Eq. (\ref{ww}) with b$^1_C=$b$^0_3$ and
b$^1_L=$b$^0_L$ will imply $M_H\simeq 3M_Z$, a very small
value for the right-handed weak current. Other gauge structures for G$_H$
which do not contain flavor changing neutral currents such as
SU(3)$_c\otimes$SU(2)$_L\otimes$[U(1)$_{y_i}]^n$ are not allowed by the
set of Higgs fields described in Sec.2; so, with the
minimal set of Higgs fields presented, G$_H$ will contain flavor changing
neutral currents and M$_H\geq 100$ TeVs and the first two terms
of the right-hand side of Eq.(\ref{ww}) will have a lower bound of 0.36.
Then if we want to reproduce the experimental value for sin$^2\theta_W(M_Z)$,
$(b^1_C-b^1_L)<0$. As
mentioned in Ref.  \cite{pz} this is a very stringent constraint which is not
satisfied by the set of Higgs fields and VeVs presented in Sec. 2.2 (in
combination with the extended survival hypothesis).
So, up to this point there is a demand for changing the minimum set of
HF and/or VEVs presented.

Now, what is the minimum change in the set of HF and/or VEVs which
properly breaks the symmetry, respect the survival hypothesis,
produces appropriate values for sin$^2\theta_W(M_Z)$, and holds the mass
hierarchy M$>>$M$_H>>$M$_Z\sim 10^2$ GeVs? The study of table I shows that
the three stages gauge hierarchy with
G$_H=$SU(6)$_L\otimes$SU(4)$_C\otimes$U(1)$_Y\otimes...$ produces
consistent results as far as we do the following two things:\\
1-Add a new set of Higgs Fields
\begin{equation}
\phi^{\prime}_{2}=\phi^{\prime}(1323)=
\phi_{2,\{a,b\}}^{\prime,\{A,B\}}+\phi^{\prime,\{\alpha,\beta\}}_{2,\{A,B\}}
+\phi^{\prime,\{a,b\}}_{2,\{\alpha,\beta\}}
\end{equation}

\noindent
with VEVs in the directions \{a,b\}=\{3,6\}=$-$\{4,5\}, \{A,B\} similar to
\{a,b\} and \{$\alpha,\beta$\}=\{5,5\};\\
2-Orient the VEVs such that
$\langle\phi^{\prime,\{a,b\}}_{2,\{\alpha,\beta\}=\{5,5\}}\rangle=
\langle\phi^{\prime,\{A,B\}}_{2,\{a,b\}}\rangle=
\langle\phi^{\{A,B\}}_{2,\{a,b\}}\rangle=0$

For this particular choice of VEVs we have that
$b^1_C=(\frac{88}{3}-\frac{2\times 12}{3}-\frac{148}{3})/4\pi$,
$b^1_L=(\frac{132}{3}-\frac{2\times 12}{3}-\frac{107}{3})/4\pi$ and
$b^1_Y=-(\frac{2\times 12}{3}+\frac{9}{14})/4\pi$, where the extended
survival hypothesis \cite{sh} was taken into account for the
contribution of the HF.

As can be seen, the Higgs fields play a fundamental role in equations
(\ref{w}) and (\ref{ww}) (the same is true for other models \cite{pal}
such as SU(15) and SU(16)). Notice also that
we achieve the stated hierarchy for M=$\langle\phi_3\rangle$,
M$_H=\langle\phi_1\rangle+\langle\phi_2\rangle+\langle\phi_2^{\prime}\rangle$,
M$_Z=\langle\phi_4\rangle$,
and that $\phi_3$ plays no role in
the evolution of the gauge coupling constants.

Plugging in the different beta functions in (\ref{w}) and (\ref{ww}) and
using the experimental values for sin$^2\theta_W(M_Z)$,
$\alpha_{EM}(M_Z)$, $\alpha_3(M_Z)$ we get the equations
\begin{displaymath}
1.10=1.02\ln (\frac{M_H}{M_Z})-2.26\ln (\frac{M}{M_H})
\end{displaymath}
\begin{displaymath}
7.83=1.25\ln (\frac{M_H}{M_Z}) -1.82\ln (\frac{M}{M_H}),
\end{displaymath}
which for M$_Z=91$ GeVs have the solutions
M$_{H}\sim  10^9$ GeVs and M$\sim  10^{12}$ GeVs.
These results are in good
agreement with the values calculated from the analysis of the
generational see-saw mechanism done in the context of
this model \cite{pzg}.

\section{Stability of the Proton}
\subsection{Baryon number for the particles}
The elementary particles in the model are the ones associated to the 105
GF, the 108 Weyl fields in $\psi(108)_L$ and the 4104 HF in
$\phi_i,i=$1-4 and $\phi^{\prime}_2$. Now, all the
elementary particles in our model
have well defined Baryon numbers. Let us see: \\
1-\underline{The GF.}\\
The 70 GF associated with SU(6)$_L\otimes$SU(6)$_R$ have BN zero
(all of them are color singlets).\\
For SU(6)$_C$ we have that the 9 leptoquarks have BN equal to 1/3, the 9
leptoquarks charge conjugated to the previous ones have BN equal
to $-1/3$ and the other 17 GF have BN equal to zero (including the 8 gluon
fields).\\
2-\underline{The Weyl Fermion Fields.}\\
All the quark fields in $\psi(\bar{6},6,1)_L$ have BN equal to 1/3,
the quark fields in $\psi(1,\bar{6},6)_L$ have BN
equal to $-1/3$ and all the other fields in $\psi(108)_L$ have BN equal
to zero.\\
3-\underline{The HF.}\\
The BN for the 4104 HF of the model are given in Table 2.

\subsection{Baryon number as a symmetry of the model}
In the subspace defined by SU(6)$_C$ the BN can be associated
with the $6\times 6$ diagonal matrix B$ = Dg.(1/3,1/3,1/3,0,0,0)$. This
matrix does not correspond to a generator of SU(6)$_C$ neither of G.

Now, the full Lagrangian
${\cal L}={\cal L}_{gauge}+{\cal L}_{Higgs} +{\cal L}_{Yukawa}$
has a U($\theta$) global symmetry given by

\begin{equation}
{\rm U}(\theta)=e^{i\theta\chi_h}
\end{equation}

\noindent
where $\chi_h$ is a constant value related to
the irreducible representation of
SU(6)$_C$ under which U($\theta$) acts (for example $\chi=1$ for
$\psi(\bar{6},6,1)$, $\chi=0$ for G(1,35,1), $\chi=-2$ for
$\phi(1,\overline{15},15)$, etc.).

Associated to U($\theta$) there is a global U(1) generator which we may
write in the subspace of SU(6)$_C$ as:
\begin{equation}
{\rm U}(1)=Dg.(1,1,1,1,1,1)/\sqrt{12}
\end{equation}
which is not an element of the Lie algebra of G either
(the value $\sqrt{12}$ is introduced just for convenience).

On the other hand, in the Lie algebra of G there is a generator,
element of the SU(6)$_C$ subalgebra, of the form
\begin{equation}
B^{\prime}=Dg.(1,1,1,-1,-1,-1)/\sqrt{12}
\end{equation}
which distinguish between quarks and leptons in the context of our model.
Therefore B can be written as

\begin{equation}
{\rm B}=[{\rm U}(1)+B^{\prime}]/\sqrt{3}
\end{equation}

\subsection{Stability of the proton}
Since the elementary particles of this model have well-defined values of BN it
is obvious that in the unbroken theory the exchange of particles cannot
break BN. This statement is also true after breaking the symmetry due
to the following two facts:

\begin{itemize}
\item Baryon number is not gauged
(there is not gauge boson associated to B).
\item $\phi_i, i$=1-4 and $\phi^{\prime}_2$ with the VEVs as stated
do not break spontaneously B. That is
${\rm B}\langle\phi_i\rangle ={\rm B}\langle\phi_2^{\prime}\rangle =0$,
$i=$1,2,3,4.
\end{itemize}
So, {\bf B is conserved in our model} \cite{florez}. Once {\bf B}
is conserved, the proton will be stable against all decays, except possible
topological effects.

Now, the single Goldstone boson associated with the broken orthogonal
combination  is eaten by the massive gauge field associated with B$^{\prime}$,
so
there are no physical Goldstone bosons and no long range force. This
mechanism in which a global symmetry emerges from the simultaneous
breaking of a gauge and global symmetry is due to t'Hooft \cite{thooft}
and was implemented in the context of GUMs in Ref. \cite{langa}.

Finally we want to mention that even
though the baryon number is conserved
in the context of this model, the lepton number is violated due to the
fact that the GF associated with U(1)$_{Y_{(B-L)}}$ generator of the
SU(6)$_C$ subalgebra, and given by \cite{pz}:
\begin{equation}
{\rm U(1)}_{Y_{(B-L)}}=\sqrt{\frac{3}{20}}Dg.(1/3,1/3,1/3,-1,1,-1)
\end{equation}
is gauged in the context of our model. So, neither L, (B$-$L) or (B+L)
are conserved quantities in the context of the model presented here.

\noindent
{\bf ACKNOWLEDGMENTS}\\
\noindent
This work was partially supported by CONACyT in M\'exico and COLCIENCIAS in
Colombia. Two of us (W.A.P. and A.Z.) thank Gordon Kane for an usefull
conversation.

\pagebreak

\begin{center}
TABLE 1.\\
Contribution of the HF to the index value for
SU(4)$_C$ and SU(6)$_L$.
\end{center}

\vspace{0.5cm}

\begin{tabular}{||c|c|c|c|c||}   \hline\hline
$\langle\phi\rangle$ & SU(4)$_C$ & C$^1_C(scalars)$ & SU(6)$_L$ &
C$^1_L(scalars)$ \\ \hline\hline
$\langle\phi_{1,[a,b]}^{[A,B]}\rangle$
&   & 0 & fourteen {\bf 15} & 56 \\ \hline
$\langle\phi^{[\alpha,\beta]}_{1,[A,B]}\rangle$
& fourteen {\bf 4} & 14 &   & 0 \\ \hline
$\langle\phi^{[a,b]}_{1,[\alpha,\beta]}\rangle$
& fifteen {\bf 4} & 15 & four {\bf 15} & 16 \\ \hline
$\langle\phi_{2,\{a,b\}}^{\{A,B\}}\rangle$
&   & 0 & fourteen {\bf 21} & 112 \\ \hline
$\langle\phi^{\{\alpha,\beta\}=\{4,5\}}_{2,\{A,B\}}\rangle$
& fourteen {\bf 4} & 14 &   & 0 \\ \hline
$\langle\phi^{\{a,b\}}_{2,\{\alpha,\beta\}\{4,5\}}\rangle$
& twenty-one {\bf 4} & 21 &  four {\bf 21} & 32 \\ \hline
$\langle\phi_{2,\{a,b\}}^{\prime\{A,B\}}\rangle$
&   & 0 & fourteen {\bf 21} & 112 \\ \hline
$\langle\phi^{\prime\{\alpha,\beta\}=\{5,5\}}_{2,\{A,B\}}\rangle$
& fourteen {\bf 10} & 84 &   & 0 \\ \hline
$\langle\phi^{\prime\{a,b\}}_{2,\{\alpha,\beta\}=\{5,5\}}\rangle$
& twenty-one {\bf 10} & 126 &  ten {\bf 21} & 80 \\ \hline
$\langle\phi^a_{4,A}\rangle$
&   & 0 & three {\bf 6} & 3 \\ \hline\hline
\end{tabular}

\pagebreak

\pagebreak

\begin{center}
TABLE 2. \\
BN for the 4104 HF
\end{center}
\renewcommand{\baselinestretch}{1.2}
\vspace{0.5cm}

\begin{tabular}{||c|c|c|c|c||}   \hline\hline
$\phi$ & $\alpha ,\beta$ & $a,b$ & $A,B$ & BN \\ \hline\hline
$\phi_{1(3),[a,b]}^{[A,B]}$ &  & $a,b=1,...,6$ & $A,B=1,...,6$ & 0 \\ \hline
$\phi^{[\alpha,\beta]}_{1(3),[A,B]}$ & $\alpha ,\beta$ =1,2,3 & &
$A,B=1,...,6$ & $-1/3$ \\
 & $\alpha ,\beta$ =4,5,6 &  & $A,B=1,...,6$ & 0 \\
 & $\alpha =1,2,3;\beta$ =4,5,6 &  & $A,B=1,...,6$ & 1/3 \\ \hline
$\phi^{[a,b]}_{1(3),[\alpha,\beta]}$ & $\alpha ,\beta$ =1,2,3 &
$a,b=1,...,6$ &  & 1/3 \\
 &$ \alpha , \beta$ =4,5,6 & $a,b=1,...,6$ &  & 0 \\
 & $\alpha =1,2,3;\beta$ =4,5,6 & $a,b=1,...,6$ &  & $-1/3$ \\ \hline
$\phi_{2,\{a,b\}}^{(\prime)\{A,B\}}$ &  & $a,b=1,...,6$ & $A,B=1,...,6$ & 0 \\
\
   hline
$\phi^{(\prime)\{\alpha,\beta\}}_{2,\{A,B\}}$ & $\alpha ,\beta$ =1,2,3 & &
$A,B=1,...,6$ & 2/3 \\
 & $\alpha ,\beta$ =4,5,6 &  & $A,B=1,...,6$ & 0 \\
 & $\alpha =1,2,3;\beta$ =4,5,6 &  & $A,B=1,...,6$ & 1/3 \\ \hline
$\phi^{(\prime)\{a,b\}}_{2,\{\alpha,\beta\}}$ & $\alpha ,\beta$ =1,2,3 &
$a,b=1,...,6$ &  & $-2/3$ \\
 &$ \alpha , \beta$ =4,5,6 & $a,b=1,...,6$ &  & 0 \\
 & $\alpha =1,2,3;\beta$ =4,5,6 & $a,b=1,...,6$ &  & $-1/3$ \\ \hline

$\phi^a_{4,A}$ &  & $a=1,...,6$ & $A=1,...6$ & 0 \\ \hline
$\phi^\alpha_{4,a}$ & $\alpha =1,2,3$ & $a=1,...,6$ &  & 1/3 \\
  & $\alpha =4,5,6$ & $ a=1,...,6$ &  & 0 \\ \hline
$\phi^A_{4,\alpha}$ & $\alpha=1,2,3$ &  & $A=1,...,6$  & $-1/3$ \\
 &  $\alpha =4,5,6$ &  & $A=1,...,6$ & 0 \\ \hline\hline
\end{tabular}

\pagebreak
\renewcommand{\baselinestretch}{2}


\begin{thebibliography}{99}

\bibitem{gmpz}
A.H.Galeana, R.Martinez, W.A.Ponce, and A.Zepeda; Phys. Rev.
{\bf D44}, 2166(1991).

\bibitem{pz}
W.A.Ponce, and A.Zepeda; Phys. Rev. {\bf D47}, xxxx (1993).
\bibitem{zee}
F.Wilczek and A. Zee; Phys. Rev. {\bf D25}, 553(1982).\\
F.Senjanovic, F.Wilczek and A.Zee; Phys. Lett. B141, 389 (1984).\\
J.Bager and S. Dimopoulos; Nucl. Phys. {\bf B244}, 247(1984).\\
D.Chang, and R. Mohapatra; Phys. Lett. {\bf 158B}, 323(1985).\\
J.Bagger {\it et al}; Nucl. Phys. {\bf B258}, 565(1985).

\bibitem{bars}
I.Bars and M. Gunaydin; Phys. Rev. Lett. {\bf 45}, 859(1980).\\
S.M.Barr; Phys. Rev. {\bf D37}, 204(1988).

\bibitem{sh}
H.Georgi; Nucl. Phys. {\bf B156}, 126 (1979).\\
R.Barbiery and D.V.Nanopoulos, Phys. Lett. {\bf 91B}, 369 (1980).

\bibitem{esh}
F. Del Aguila, and L.Iba\~nez; Nucl. Phys. {\bf B177}, 60 (1981).\\
H.Georgi and S.Dimopoulos, Phys. Lett. {\bf 140B}, 67 (1984).

\bibitem{apel}
T.Appelquist and J.Carazzone; Phys. Rev. {\bf D11}, 2856 (1975).

\bibitem{amaldi}
U. Amaldi {\it et al.}; Phys. Lett. {\bf B281}, 374(1992).

\bibitem{pal}
B.Brahmachari {\it et al}; Phys. Rev. {\bf D45}, 2467 (1993).\\
N.G.Deshpande, E.Keith, and P.B.Pal; Phys. Rev. {\bf D47}, 2897 (1993).

\bibitem{pzg}
W.A.Ponce and A.Zepeda; ``{\it Generational seesaw mechanism in }
[SU(6)]$^3\times$Z$_3$". CINVESTAV preprint (1992). Submited for
publication.

\bibitem{florez}
J.B.Florez; Ph.D. Thesis, CINVESTAV 1993 (unpublished).

\bibitem{thooft}
G.t'Hooft; Nucl. Phys. {\bf B35}, 167 (1971).

\bibitem{langa}
P.Langacker, G.Segre, and A.Weldon, Phys. Lett. {\bf 73B}, 87 (1978).
{\it ibid} Phys. Rev. {\bf D18}, 552 (1978).\\
M.Gell-Mann, P.Ramond and R. Slanski; Rev. of Mod. Phys. {\bf 50},
721 (1978).

\end{thebibliography}
\end{document}